%Paper: hep-ph/9209222
%From: LONDON@lps.umontreal.ca
%Date: Wed, 9 Sep 1992 11:40:01 -0400 (EDT)

% -----------------------------------------------------------------------
% Size and shape
% -----------------------------------------------------------------------

\font\titlefont = cmr10 scaled\magstep 4
\font\sectionfont = cmr10
\font\littlefont = cmr5% for equation names in draftmode

%comment this out for tow column
\magnification = 1200% 12 point

%\global\hsize = 5in% size of text
%\global\lmargin = 0.125in%
\global\baselineskip = 1.2\baselineskip% line skip
\global\parskip = 4pt plus 0.3pt% paragraph skip
\global\abovedisplayskip = 18pt plus3pt minus9pt
\global\belowdisplayskip = 18pt plus3pt minus9pt
\global\abovedisplayshortskip = 6pt plus3pt
\global\belowdisplayshortskip = 6pt plus3pt

% -----------------------------------------------------------------------
% Draft mode stuff
% -----------------------------------------------------------------------

\def\endignore{}
\def\ignore #1\endignore{}% use to "comment out" text

\newcount\dflag% draft mode flag
\dflag = 0% initialize

% Time commands ---------------------------------------------------------

\def\monthname{\ifcase\month% for month numbers
\or Jan \or Feb \or Mar \or Apr \or May \or June%
\or July \or Aug \or Sept \or Oct \or Nov \or Dec% month name
\fi}

\def\timestring{{\count0 = \time%
\divide\count0 by 60%
\count2 = \count0% hour
\count4 = \time%
\multiply\count0 by 60%
\advance\count4 by -\count0% minutes
\ifnum\count4 < 10 \toks1 = {0}% leading 0 for minutes
\else \toks1 = {} \fi%
\ifnum\count2 < 12 \toks0 = {a.m.}% before noon
\else \toks0 = {p.m.}% after noon
\advance\count2 by -12%
\fi%
\ifnum\count2 = 0 \count2 = 12 \fi% make midnight = 12
\number\count2 : \the\toks1 \number\count4%
\thinspace \the\toks0}}

% Draft mode commands ---------------------------------------------------

% -----------------------------------------------------------------------
% Headers
% -----------------------------------------------------------------------

\def\endtitle{}
\def\title#1\endtitle{\vskip.5in\titlefont
\global\baselineskip = 2\baselineskip% set line skip
#1\vskip.4in% title
\baselineskip = 0.5\baselineskip\rm}

\def\endauthors{}
\def\authors#1\endauthors{#1}

\def\endabstract{}
\def\abstract#1\endabstract{\vskip .3in%
\centerline{\sectionfont\bf Abstract}%
\vskip .1in
\noindent#1}

\newcount\nsection% section counter
\newcount\nsubsection% subsection counter

% start new section
\def\section#1{\global\advance\nsection by 1% increment section number
\nsubsection=0
% section title
\bigskip\noindent\centerline{\sectionfont \bf \number\nsection.\ #1}
\bigskip\rm\nobreak}% back to normal

% start new subsection
\def\subsection#1{\global\advance\nsubsection by 1% increment subsection number
% subsection title
\bigskip\noindent\sectionfont \sl \number\nsection.\number\nsubsection)\
#1\bigskip\rm\nobreak}% back to normal

% unnumbered itemized topics

% start new appendix
\def\appendix#1#2{\bigskip\noindent%
\centerline{\sectionfont \bf Appendix #1.\ #2}% appendix title
\bigskip\rm\nobreak}% back to normal

% -----------------------------------------------------------------------
% References
% -----------------------------------------------------------------------

\newcount\nref% create counter for references
\global\nref = 1% initialize it

\def\ref#1#2{\xdef #1{[\number\nref]}% define reference label
\ifnum\nref = 1\global\xdef\therefs{\noindent[\number\nref] #2\ }% first ref
\else% not the first ref
\global\xdef\oldrefs{\therefs}% old reference list
\global\xdef\therefs{\oldrefs\vskip.1in\noindent[\number\nref] #2\ }%
\fi%
\global\advance\nref by 1% advance label
}

\def\listrefs{\vfill\eject\section{References}\therefs}

% -----------------------------------------------------------------------
% Equations
% -----------------------------------------------------------------------

\newcount\cflag% create custom flag
\newcount\nequation% create equation counter
\global\nequation = 1% initialize it
\def\eqlabel{(1)}% initialize equation label

% Increment equation counter
\def\nexteqno{\ifnum\cflag = 0% if no custom numbering
\global\advance\nequation by 1% advance number
\fi% end of conditional
\global\cflag = 0% reset custom flag
\xdef\eqlabel{(\number\nequation)}}% define equation label

% Decrement equation counter
\def\lasteqno{\global\advance\nequation by -1% decrease number
\xdef\eqlabel{(\number\nequation)}}% define equation label

% Label equation
\def\label#1{\xdef #1{(\number\nequation)}% define equation name
\ifnum\dflag = 1% if in draft mode
{\escapechar = -1% locally remove "\"
\xdef\draftname{\littlefont\string#1}}% define draft name
\fi}

% Custom label equation
\def\clabel#1#2{\xdef\eqlabel{(\number\nequation #2)}% define custom label
\global\cflag = 1% set custom flag
\xdef #1{\eqlabel}% label equation
\ifnum\dflag = 1% if in draft mode
{\escapechar = -1% locally remove "\"
\xdef\draftname{\string#1}}% define draft name
\fi}

% Completely custom label equation
\def\cclabel#1#2{\xdef\eqlabel{#2)}% define custom label
\global\cflag = 1% set custom flag
\xdef #1{\eqlabel}% label equation
\ifnum\dflag = 1% if in draft mode
{\escapechar = -1% locally remove "\"
\xdef\draftname{\string#1}}% define draft name
\fi}

% Display equation stuff ------------------------------------------------

% End of equation
\def\eeq{}

% Begin displayed unnumbered equation
\def\eqnn #1\eeq{$$ #1 $$}

% Begin displayed numbered equation
\def\eq #1\eeq{\xdef\draftname{\ }% default = no draft name
$$ #1% print equation
\eqno{\eqlabel \rlap{\ \draftname}} $$% print equation number
\nexteqno}% increment equation number

% Print equation number

% Equation array stuff --------------------------------------------------

% End line with equation number
% clear draft name

% Last eol
\def\eeol{& \eqlabel \rlap{\ \draftname}% print equation number
\nexteqno% increment equation number
\xdef\draftname{\ }}% clear draft name

% End line without equation number
\def\eolnn{\cr% end of line
\global\cflag = 0% reset custom flag  (just to be sure)
\xdef\draftname{\ }}% clear draft name

% Last eol without equation number
% clear draft name

% begin equation array
\def\eqa #1\eeq{\xdef\draftname{\ }% clear draft name
$$ \eqalignno{ #1 } $$% print equation
\global\cflag = 0}% reset custom flag

% -----------------------------------------------------------------------
% Useful abbreviations
% -----------------------------------------------------------------------

% -----------------------------------------------------------------------
% Journal abbreviations
% -----------------------------------------------------------------------

\def\npb#1#2#3{{\it Nucl. Phys.} {\bf B#1} (19#2) #3}

\def\prd#1#2#3{{\it Phys. Rev.} {\bf D#1} (19#2) #3}
\def\pr#1#2#3{{\it Phys. Rev.} {\bf #1} (19#2) #3}

\def\prl#1#2#3{{\it Phys. Rev. Lett.} {\bf #1} (19#2) #3}

\def\zpc#1#2#3{{\it Zeit. Phys.} {\bf C#1} (19#2) #3}

% Math parameters -----------------------------------------------------------

\global\nulldelimiterspace = 0pt

% Math relations ------------------------------------------------------------

% Math operations -----------------------------------------------------------

\def\frac#1#2{{{#1} \over {#2}}\,}  % fraction

  % small fraction

  % derivative
  % partial derivative

\def\Dsl{\hbox{/\kern-.6000em\it D}} % D slash
\def\dsl{\hbox{/\kern-.5600em$\partial$}}
\def\pxpsl{\hbox{/\kern-.5600em$p$}}
\def\ssl{\hbox{/\kern-.5600em$s$}}
\def\epssl{\hbox{/\kern-.5600em$\epsilon$}}
\def\delsl{\hbox{/\kern-.7000em$\nabla$}}
\def\lxpsl{\hbox{/\kern-.5600em$l$}}
\def\kxpsl{\hbox{/\kern-.5600em$k$}}
\def\qxpsl{\hbox{/\kern-.5600em$q$}}
\def\sla#1{\raise.15ex\hbox{$/$}\kern-.57em #1}% Feynman slash

% Math accents --------------------------------------------------------------

%\def\supsub#1#2{\mathstrut^{\vphantom{\dagger}#1}_{\vphantom{A}#2}}
%\def\sub#1{\mathstrut^{\vphantom{\dagger}}_{\vphantom{A}#1}}
%\def\sup#1{\mathstrut_{\vphantom{A}}^{\vphantom{\dagger}#1}}
%\def\rsub#1{\mathstrut^{\vphantom{\dagger}}_{\vphantom{A}\rm #1}}
%\def\rsup#1{\mathstrut_{\vphantom{A}}^{\vphantom{\dagger}\rm #1}}

\def\roughly#1{\mathrel{\raise.3ex\hbox{$#1$\kern-.75em\lower1ex\hbox{$\sim$}}}}

% Alphabets ------------------------------------------------------------------

% Lower Case Bold Face

% Upper Case Bold Face

\def\Bfw{{\bf W}}

% Upper Case Script

\def\Sca{{\cal A}}

\def\Scd{{\cal D}}

\def\Scl{{\cal L}}

\def\Scw{{\cal W}}

\def\Scz{{\cal Z}}

% Math functions -------------------------------------------------------------

\def\tr{\mathop{\rm tr}}

% Math constructs ------------------------------------------------------------

% bras 'n' kets

% integral measures

% Abbreviations --------------------------------------------------------------

% units

% References -----------------------------------------------------------

\ref\LWWV{The most general trilinear gauge boson couplings are discussed in
K.J.F. Gaemers and G.J. Gounaris, \zpc{1}{79}{259};
K. Hagiwara, R.D. Peccei, D. Zeppenfeld and K. Hikasa, \npb{282}{87}{253}.}

\ref\gbloop{For references to the literature, see Refs.~5,7.}

\ref\stu{M.E. Peskin and T. Takeuchi, \prl{65}{90}{964};
W.J. Marciano and J.L. Rosner, \prl{65}{90}{2963};
D.C. Kennedy and P. Langacker, \prl{65}{90}{2967}; \prd{44}{91}{1591}.}

\ref\tlimit{P. Langacker, U. Penn preprint UPR-0492T (1991).}

\ref\derujula{A. de R\'ujula, M.B. Gavela, P. Hernandez and E. Mass\'o,
CERN preprint CERN-Th.6272/91, 1991.}

\ref\equivalence{M.S. Chanowitz, M. Golden and H. Georgi,
\prd{36}{87}{1490}; C.P. Burgess and David London, preprint McGill-92/04,
UdeM-LPN-TH-83, 1992.}

\ref\cutoff{C.P. Burgess and David London, preprint McGill-92/05,
UdeM-LPN-TH-84, 1992.}

\ref\weinberg{S. Weinberg, \pr{140}{65}{B516}; in {\it Asymptotic Realms Of
Physics} (Cambridge, 1981).}

\ref\wrel{For an explicit example, see the appendix of Ref.~7.}

% Local Definitions ---------------------------------------------------------

\def\smgroup{SU(2)_L\times U(1)_Y}

% Title page ----------------------------------------------------------------

\rightline{September 1992}
\rightline{UdeM-LPN-TH-105}
\rightline{McGill-92/39}

\title
\centerline{Loops, Cutoffs and}
\centerline{Anomalous Gauge Boson Couplings\footnote{$^\dagger$}
{\baselineskip=0.5\baselineskip\rm Invited talk presented by David London
at the {\it XXVI International Conference on High Energy Physics}, Dallas,
USA, August 1992}\baselineskip=2.0\baselineskip}
\endtitle

\authors
\centerline{David London${}^a$ and C.P. Burgess${}^b$}
\vskip .15in
\centerline{\it ${}^a$ Laboratoire de Physique Nucl\'eaire, Universit\'e de
Montr\'eal}
\centerline{\it C.P. 6128, Montr\'eal, Qu\'ebec, CANADA, H3C 3J7.}
\vskip .1in
\centerline{\it ${}^b$ Physics Department, McGill University}
\centerline{\it 3600 University St., Montr\'eal, Qu\'ebec, CANADA, H3A 2T8.}
\endauthors

\abstract
We discuss several issues regarding analyses which use loop calculations to
put constraints on anomalous trilinear gauge boson couplings (TGC's). Many
such analyses give far too stringent bounds. This is independent of
questions of gauge invariance, contrary to the recent claims of de R\'ujula
et. al., since the lagrangians used in these calculations {\it are} gauge
invariant, but the $\smgroup$ symmetry is nonlinearly realized. The real
source of the problem is the incorrect use of cutoffs -- the cutoff
dependence of a loop integral does not necessarily reflect the true
dependence on the heavy physics scale $M$. If done carefully, one finds
that the constraints on anomalous TGC's are much weaker. We also compare
effective lagrangians in which $\smgroup$ is realized linearly and
nonlinearly, and discuss the role of custodial $SU(2)$ in each formulation.
\endabstract
\vfill\eject

% Main text ----------------------------------------------------------------

Although the standard model of the weak and electromagnetic interactions
has been extremely successful in explaining all experimental results to
date, the gauge structure of the theory has not yet been tested. This task
will be accomplished in the coming years in such experiments as LEP200 and
the TeVatron, where the three-gauge-boson self couplings will be directly
measured. Of course, it is hoped that new physics will be observed. With
this in mind, these experiments will search for, among other things,
anomalous trilinear gauge boson couplings (TGC's) not found in the standard
model.

If one wants to parametrize new physics such as anomalous TGC's, the
easiest way to do so is to use an effective lagrangian. In order to define
the low-energy effective lagrangian which parametrizes the new physics, it
is necessary only to specify the particle content and the symmetries of the
low-energy theory. In dealing with anomalous TGC's, one basically has three
choices:
\item{1.} Linearly Realized $\smgroup$: Here, symmetry breaking is
accomplished through the Higgs mechanism, and the low-energy theory
includes the standard model Higgs doublet.
\item{2.} Nonlinearly Realized $\smgroup$: In these effective lagrangians,
the symmetry breaking mechanism is unspecified -- the low-energy theory
contains only those pseudo-Nambu-Goldstone bosons which are eaten to give
mass to the $W$'s and $Z$'s. These are also known as ``chiral
lagrangians''.
\item{3.} Only $U(1)_{em}$ Gauge Invariance: In this case, the low-energy
effective lagrangian is required only to obey electromagnetic gauge
invariance. Massive $W$'s and $Z$'s and their interactions are simply put
in by hand.

At this point we would like to make a comment regarding the first choice
above. It is, in fact, a very strong assumption to assume that, even in the
presence of new physics which gives rise to anomalous TGC's, $\smgroup$ is
still broken to $U(1)_{em}$ via a single Higgs doublet. After all, given
that there is new physics, the method of symmetry breaking might be quite
different from that of the standard model. Furthermore, anomalous TGC's
often involve the longitudinal components of the gauge bosons, which are
intimately connected to the symmetry breaking mechanism. Most conclusions
based upon linearly realized $\smgroup$ could be altered if one changes,
for example, the Higgs content. Of course, this does not imply that it is
incorrect to use an effective lagrangian based on linearly realized
$\smgroup$. However, it is more conservative to use nonlinearly realized
$\smgroup$. We will briefly return later to a comparison of the two
formulations.

Up to now, most analyses which concern themselves with anomalous TGC's use
the third option -- the effective lagrangian obeys only electromagnetic
gauge invariance. That is,
\eqa
\label\wwvlag
\Scl \sim & ~M_W^2 W_\mu^\dagger W^\mu + {1\over 2} M_Z^2 Z_\mu Z^\mu
+ i \kappa_V W_\mu^\dagger W_\nu V^{\mu\nu} \eolnn
&~ + i {\lambda_V \over M^2}
W_{\lambda\mu}^\dagger W^\mu_{~~\nu} V^{\nu\lambda}
- g_4^Z W_\mu^\dagger W_\nu \left( \Scd^\mu Z^\nu + \Scd^\nu Z^\mu\right)
+ ... \eeol
\eeq
Here, we have written only a subset of the possible terms, including
explicit masses for the $W$ and $Z$, and several triple gauge boson
vertices \LWWV. In the above, $V^\mu$ represents either the photon or the
$Z$, $W^\mu$ is the $W^-$ field, $\Scd_\mu$ is the electromagnetic
covariant  derivative, and $W_{\mu\nu}=\Scd_\mu W_\nu - \Scd_\nu W_\mu$
(and similarly for $V_{\mu\nu}$). The $g_4^Z$ term is a CP violating TGC
called (for obscure reasons) the anapole coupling, which we will use in our
examples below.

Now, given that these anomalous three-gauge-boson vertices will not be
measured for several years, it is reasonable to ask whether limits can be
placed on them using current data. Obviously the only constraints can come
through contributions to loop-induced processes, and a number of papers
have considered this possibility \gbloop. Let us illustrate a typical such
calculation.

Consider the CP violating anapole coupling $g_4^Z$ defined in Eq.~\wwvlag\
above. This will contribute at one loop to the $W$- and $Z$-masses.
However, because it is CP violating, there will be a nonzero result only if
this anomalous TGC appears at both vertices in the loop diagrams, as in
Fig.~2. Now, the anapole coupling is nonrenormalizable, and hence the
diagrams diverge. The standard way to regularize the divergent loop
integrals is to simply insert a cutoff. If this is done, then one finds
(keeping only the highest divergence)
\eqa
\delta\pi_{WW} \left(q^2\right) = &
{}~- { \left(g_4^Z\right)^2 \over 6 \pi^2 }
\thinspace { \Lambda^6 \over M_W^2 M_Z^2 } \eolnn
\delta\pi_{ZZ} \left(q^2\right) = &
{}~- { \left(g_4^Z\right)^2 \over 144 \pi^2 }
\thinspace { \Lambda^4 \over M_W^4 } \thinspace q^2. \eeol
\eeq
Since the values of $\delta\pi_{WW}$ and $\delta\pi_{ZZ}$ at $q^2=0$ are
unequal, there will be a nonzero contribution to the deviation of the
$\rho$-parameter from unity, often parametrized using the $T$-parameter of
Peskin and Takeuchi \stu. Taking $\Lambda$ to be the scale of new physics,
say 1 TeV, and using the present limit of $\vert T\vert<0.8$ \tlimit, one
obtains a very stringent constraint on $g_4^Z$:
\eq
\label\overestimate
g_4^Z < 3.5 \times 10^{-4} \left( { 1~{\rm TeV} \over \Lambda } \right)^3.
\eeq

Intuitively, this result seems suspect. After all, the anomalous coupling
contributes only at one loop, and the quantity which is used to constrain
$g_4^Z$, the $\rho$-parameter, has only been measured to a precision of a
couple of percent or so. It's not as if this limit comes from an extremely
well-measured process such as $\mu\to e\gamma$, for example.

In fact, there is a recent paper by de R\'ujula et.~al.~\derujula, in
which they claim that this constraint is a considerable overestimate (and
similarly for calculations which predict large measurable effects in
loop-induced processes involving anomalous TGC's). The reason, they say, is
that the lagrangian used (Eq.~\wwvlag) is not $\smgroup$ gauge invariant.

This claim is only partly correct. It is true that the result in
Eq.~\overestimate\ is an overestimate, for reasons we will explain later.
However, the reason has nothing whatsoever to do with gauge invariance. The
point is that the lagrangian in Eq.~\wwvlag, which obeys only
electromagnetic gauge invariance, is equivalent, term by term, to a chiral
lagrangian in which $\smgroup$ is present but nonlinearly realized
\equivalence.

Briefly, the proof goes as follows. To construct a lagrangian which
contains a nonlinear realization of $\smgroup$, broken to $U(1)_{em}$, one
introduces the matrix-valued scalar field $\xi(x)\equiv exp\left[i
X_a\phi^a/f\right]$, in which the $\phi^a$ are the Nambu-Goldstone bosons,
and the $X_a$ are the broken generators. (The afficionados may remark that,
as we have written it, $\xi(x)$ respects custodial $SU(2)$, since there is a
common decay constant $f$ for each of the $\phi^a$'s. However, we could
equally well have ignored this symmetry by writing $f_a$ -- the proof is
independent of the existence of a custodial $SU(2)$.) With $\xi(x)$, one
can define a nonlinear ``covariant derivative''
\eq
\Scd_\mu(\xi) \equiv \xi^\dagger \partial_\mu\xi -i \xi^\dagger \Bfw_\mu
\xi,
\eeq
in which $\Bfw_\mu = g_2 W^a_\mu \, T_a + g_1 B_\mu \, Y$. In terms of
$\Scd_\mu(\xi)$, one constructs the three fields
\eqa
e\, \Sca_\mu & \equiv i \, \tr[ Q \Scd_\mu(\xi)], \eolnn
\sqrt{g_1^2 + g_2^2} \; \Scz_\mu &
           \equiv 2i \, \tr[(T_3 - Y) \Scd_\mu(\xi)],\eolnn
g_2 \, \Scw^\pm_\mu & \equiv i\sqrt{2} \, \tr[T_\mp \Scd_\mu(\xi)]. \eeol
\eeq
The significance of these three quantities, $\Sca_\mu$, $\Scz_\mu$ and
$\Scw_\mu$ is that, under arbitrary $\smgroup$ transformations, they
transform purely electromagnetically. Thus, any lagrangian which is
constructed using these quantities and which is required to obey
electromagnetic gauge invariance will automatically obey the full
$\smgroup$ symmetry. In unitary gauge, these fields reduce to the standard
photon, $W$ and $Z$ fields:
\eq
\Sca_\mu \leftrightarrow A_\mu, \quad
\Scz_\mu \leftrightarrow  Z_\mu, \quad
\Scw_\mu^\pm \leftrightarrow W^\pm_\mu.
\eeq
With this construction one sees explicitly that $\smgroup$ gauge invariance
is {\it automatic} for any lagrangian which obeys $U(1)_{em}$ gauge
invariance. Therefore, the claim that Eq.~\wwvlag\ is not gauge invariant
is simply a red herring.

The real reason that Eq.~\overestimate, and other results like it, are
incorrect is due to the improper use of cutoffs \cutoff. This can be seen
as follows. Suppose we knew what the full theory was at the new physics
scale, $M$, which is much larger than $m$, the scale of the light physics.
Now let us calculate the effect of the heavy physics on a light particle
mass such as $M_W$, for example. The contribution one obtains after
integrating out the heavy particles has the following form:
\eq
\label\fullresult
\delta\mu^2 (m,M) = a M^2 + b m^2 + c \frac{m^4}{M^2} + \cdots
\eeq
The dots represent terms which are of higher order in the small mass
ratio $m^2/M^2$, and the coefficients $a$, $b$, $c$, ... may depend at most
logarithmically on this ratio. Note that there is no term of the form
$M^4/m^2$. There is an old paper of Weinberg \weinberg\ which explains that
such terms are not allowed since only logarithmic infrared divergences are
allowed at zero temperature in four spacetime dimensions.

Now suppose we split this calculation up into a ``high-energy'' piece and a
``low-energy'' piece by choosing a cutoff $\Lambda$ such that
$M\gg\Lambda\gg m$. In this case, the two contributions can have the form
\eqa
\label\heresult
\delta \mu^2_{\rm he} (m,\Lambda,M) &= a^\prime M^2 + b^\prime \Lambda^2
+ c^\prime {\Lambda^4\over m^2} + \cdots \eolnn
\delta \mu^2_{\rm le} (m,\Lambda,M) &= b^{\prime\prime} \Lambda^2
+ c^{\prime\prime} {\Lambda^4\over m^2} + \cdots \eeol
\eeq
Obviously, this is just a reorganization of the full calculation
(Eq.~\fullresult), so we must have
\eq
\delta \mu^2(m,M) = \delta \mu^2_{\rm le}(m,\Lambda,M)
+ \delta \mu^2_{\rm he}(m,\Lambda,M).
\eeq
Furthermore, the full result is independent of the cutoff $\Lambda$, which
implies that
\eq
a=a^\prime ~,~~~~~~~ b^\prime=-b^{\prime\prime} ~,~~~~~~~
c^\prime=-c^{\prime\prime}~,~~~~~~~\cdots
\eeq
In other words, all quadratic and higher dependence on $\Lambda$ in the
low-energy piece of the calculation is simply cancelled by counterterms
coming from the high-energy piece of the calculation! Note also that the
coefficient $b^{\prime\prime}$ is unrelated to the coefficient $a$. That
is, a calculation of the quadratic divergence in the low-energy theory does
not tell you how the full calculation depends on $M^2$.

The point is that there is no physical significance to terms containing the
cutoff $\Lambda$. Another way of saying this, perhaps more to the point, is
that {\it cutoffs do not accurately track the true heavy mass dependence of
the full theory}.

There is one exception to this statement -- the case of a logarithmic
divergence. Suppose that, in the full theory, there were a term of the form
\eq
\delta\mu^2 \sim d m^2 \log\left({M^2\over m^2}\right).
\eeq
If one used a cutoff, the high-energy and low-energy contributions would be
\eqa
\delta \mu^2_{\rm he} & \sim d^\prime m^2
\log\left({M^2\over \Lambda^2}\right),\eolnn
\delta \mu^2_{\rm le} & \sim d^{\prime\prime} m^2
\log\left({\Lambda^2\over m^2}\right).\eeol
\eeq
Clearly, cancellation of the $\Lambda$-dependence requires
$a=a^\prime=a^{\prime\prime}$. This is the only case in which the
low-energy cutoff dependence accurately reflects the true dependence on the
heavy mass $M$.

Before returning to the example of the anapole, let us explicitly
demonstrate in a toy model the fact that the cutoff dependence found in a
low-energy loop calculation is in general unrelated to the true heavy mass
dependence. Consider a renormalizable model with only two scalars: $\psi$,
which has mass $M$, and $\phi$, of mass $m$. The potential describing the
interactions of these two scalars can be split up into two pieces,
depending on whether the potential is odd or even under separate
reflections of the fields. We have $U = U_+ + U_-$, with
\eqa
U_+ = & ~{1\over 2}m^2\phi^2 + {1\over 2}M^2\psi^2 + \lambda\phi^4 +
\lambda^\prime\psi^4 + \lambda^{\prime\prime}\phi^2\psi^2, \eolnn
U_- = & ~{1\over 3}h\psi^3 + {1\over 3}g\psi\phi^3. \eeol
\eeq
Note that, in principle, other terms could have been included in $U_-$, but
we assume that these are the only terms which appear in the lagrangian at
some scale, $\mu_0$.

Now consider integrating out the field $\psi$, so that the potential is a
function of $\phi$ only, $V(\phi)$, and consider just the $V_-$ piece of
the potential. At tree and one-loop levels, the lowest order terms which
appear are (see Fig.~2)
\eqa
{\hbox{tree level:}} & ~~~~~V_-(\phi) = {hg^3\over 81 M^6} \phi^9, \eolnn
{\hbox{one-loop level:}} & ~~~~~V_-(\phi) = -{hg\over 92\pi^2} \phi^3
\left[ 1 + 2 \log\left( {M^2\over\mu_0^2}\right) \right]. \eeol
\eeq
What is important to note here is that the dependence of the $\phi^3$ term
on the heavy mass $M$ is logarithmic.

Now suppose that the heavy physics, characterized by scale $M$, were
unknown. In this case one assumes the most general form for $V_-$,
\eq
V_-(\phi) = aM\phi^3 + {b\over M}\phi^5 + {c\over M^3}\phi^7
                     + {d\over M^5}\phi^9 + ...
\eeq
In order to make the connection with the anapole calculation we showed you
earlier, consider the contribution of the $d$-term to the $a$-term. In
other words, $aM\phi^3$ plays the role of the $\rho$-parameter, while
$(d/M^5)\phi^9$ acts like an anomalous $WWV$ coupling. Here one finds a
contribution at 3 loops (Fig.~3),
\eq
\delta a \sim \left({\Lambda^2\over 16\pi^2 M^2}\right)^3 d.
\eeq
The upshot is that one finds that the cutoff dependence of the $\phi^3$
term goes like $\Lambda^6$. However, we have already determined the true
dependence on the heavy mass to be logarithmic. This demonstrates
explicitly that cutoffs do not accurately track the true dependence of
the full theory on the heavy mass scale. This also demonstrates that the
issue of gauge invariance is indeed a red herring, since it is clear that
same type of problems arise in models which contain only scalars.

Let us now return to our original example of the anapole contribution to the
$W$- and $Z$-masses. Given the problems with cutoffs, how can one extract
physically meaningful bounds on the anapole coupling, given that there is a
nonzero contribution to the $\rho$-parameter? The easiest way to do this
(but by no means the only way) is not to use cutoffs at all to regularize
the divergent integrals. Instead, one uses dimensional regularization,
supplemented by the decoupling subtraction renormalization scheme.

Using dimensional regularization, the divergent pieces of the diagrams in
Fig.~1 are
\eqa
\label\rhodimreg
\delta\pi_{WW} \left(q^2\right)\vert_{q^2=0} = &
{}~- { \left(g_4^Z\right)^2 \over 4 \pi^2 } \thinspace {3 M_W^2 \over 2}
\left[1 + {M_Z^2\over M_W^2} - {M_Z^4\over M_W^4}\right] {2\over \epsilon},
\eolnn
\delta\pi_{ZZ} \left(q^2\right)\vert_{q^2=0} = & ~0, \eeol
\eeq
where $\epsilon=n-4$ in $n$ spacetime dimensions. The key point now is the
following. In the most general effective lagrangian, there will be a term
contributing directly to the $T$-parameter ($\Delta\rho$). The contribution
of Eq.~\rhodimreg\ to $\Delta\rho$ {\it renormalizes} this direct
contribution:
\eq
\label\mixing
\alpha T(\mu^2) = \alpha T \left({\mu'}^2 \right) - \frac{3}{8\pi^2} \;
\left( g_4^Z \left({\mu'}^2 \right) \right)^2 \;
\left[ 1 + \frac{M_Z^2}{M_W^2} - \frac{M_Z^4}{M_W^4}\right] \;
\ln \left( \frac{{\mu'}^2}{\mu^2} \right).
\eeq
This shows how the two operators in the effective lagrangian, $T$ and
$g_4^Z$, mix as the lagrangian is renormalized and evolved down from scale
$\mu^\prime$ to scale $\mu$ (in the absence of thresholds). This is a point
which seems to have been overlooked in many of the analyses which deal with
anomalous TGC's. Even in an effective (nonrenormalizable) lagrangian, the
parameters must be renormalized. In general, this requires an infinite
number of counterterms, but this is of no consequence since the effective
lagrangian already contains an infinite number of terms. Note also that
there is, in general, a contribution which depends quadratically on the
heavy mass scale $M$. This is contained in the initial condition
$T\left(M^2 \right)$, which is, however, {\it incalculable} if one does not
know the underlying theory.

If one assumes no accidental cancellation between the two terms on the
right hand side of Eq.~\mixing, one can now use the experimental limit on
$\vert T\vert$ to constrain the anapole coupling:
\eq
g_4^Z < 0.24~~~~~~~~~~({\hbox{@ 1 TeV}}).
\eeq
This is 3 orders of magnitude weaker than the bound found using cutoffs
(Eq.~\overestimate)!

Before concluding, we would like to return to a subject we briefly touched
upon at the beginning, namely the comparison of conclusions based upon
effective lagrangians with linearly and nonlinearly realized $\smgroup$.
One of the terms in Eq.~\wwvlag\ is the electric quadrupole moment of the
$W$,
\eq
i {\lambda_V \over M^2}
W_{\lambda\mu}^\dagger W^\mu_{~~\nu} V^{\nu\lambda}.
\eeq
In the literature, one often sees statements to the effect that
$\lambda_\gamma=\lambda_Z$ (modulo $\cot\theta_W$). The reasons given vary
-- occasionally gauge invariance or custodial $SU(2)$ symmetry are invoked,
and sometimes this relation is required in order to avoid too large
contributions to well-measured quantities which are in agreement with the
standard model. The fact is, none of these reasons is valid -- there is no
reason, in general, to require $\lambda_\gamma=\lambda_Z$.

There are basically two sources of confusion. First of all, if one
calculates the contribution to $\Delta\rho$ using a cutoff, one finds
\eq
\Delta\rho \sim \left(\lambda_\gamma - \lambda_Z\right)
{\Lambda^4 \over M_W^4}.
\eeq
This has led some authors to require $\lambda_\gamma=\lambda_Z$ in order to
avoid large contributions to $\Delta\rho$ for large values of the cutoff.
However, as we have argued above, this cutoff behaviour is not physically
meaningful -- this type of term is cancelled by a counterterm coming from
the high-energy theory, and no conclusions as to the relative size of
$\lambda_\gamma$ and $\lambda_Z$ should be drawn.

A more important, and subtle, source of confusion is that the relation
$\lambda_\gamma=\lambda_Z$ {\it is} true, to a good approximation, if one
uses an effective lagrangian with a linearly realized $\smgroup$ gauge
symmetry. If there is only one Higgs doublet, then it necessarily follows
that
\eq
\label\wthree
W_{3\mu} = Z_\mu \cos\theta_W + A_\mu \sin\theta_W.
\eeq
Since the standard model Higgs sector possesses an approximate custodial
$SU(2)$ symmetry, one is led quite naturally in this context to
$\lambda_\gamma=\lambda_Z$. Even if one were to add more Higgs doublets,
for example, this relation would continue to be approximately true.

On the other hand, and this is where the subtlety arises, if one realizes
the symmetry nonlinearly, then it does {\it not} necessarily follow that
$\lambda_\gamma=\lambda_Z$. Although the symmetry breaking sector continues
to possess an approximate custodial $SU(2)$ symmetry, the relation among
$W_{3\mu}$, $Z_\mu$ and $A_\mu$ need not be that found in Eq.~\wthree\
above \wrel. In general, there is no reason, neither gauge invariance nor
custodial $SU(2)$, for $\lambda_\gamma$ and $\lambda_Z$ to be related.

This is the point of this discussion. In general, an effective lagrangian
has an infinite number of terms, the coefficients of which are arbitrary
and independent. It is possible to relate some of these coefficients by
imposing certain symmetries, or to construct the effective lagrangian in a
special way. However, the lagrangian thus obtained is less general. This is
the case for an effective lagrangian with $\smgroup$ realized linearly. The
assumption of the breaking of the symmetry via the Higgs mechanism results
in certain relationships among the parameters of the low-energy effective
lagrangian. These relationships are not present if one makes no assumption
about the mechanism of symmetry breaking, i.e.~if one realizes the symmetry
nonlinearly. Again, this is not to say that the linearly realized effective
lagrangian is incorrect; however, it is more constrained than an effective
lagrangian with nonlinearly realized $\smgroup$.

To conclude,
\item{1.} Bounds on anomalous trilinear gauge boson couplings which are
obtained from their contributions to loop diagrams are often significantly
overestimated (and similarly for predictions of large effects in
loop-induced processes).
\item{2.} This is unrelated to any questions of gauge invariance. The fact
is, any lagrangian which obeys Lorentz invariance and electromagnetic gauge
invariance is equivalent, term by term, to a lagrangian in which the
$\smgroup$ symmetry is present, with the breaking $\smgroup\to U(1)_{em}$
nonlinearly realized.
\item{3.} The real source of the problem is the incorrect use of cutoffs in
estimating the effect of heavy masses in the loop diagrams. In general, the
cutoff behaviour does not properly track the true dependence on the heavy
mass scale $M$. (The one exception is the case of a logarithmic
divergence.)
\item{4.} A more straightforward way to do the calculation is not to use
cutoffs at all to regularize the divergent integrals. Instead one uses
dimensional regularization, supplemented by the decoupling subtraction
renormalization scheme, to calculate the effect of anomalous
triple-gauge-boson couplings in loops.
\item{5.} In general, there is no reason to have relationships such as
$\lambda_\gamma=\lambda_Z$ among the parameters of the low-energy effective
lagrangian -- neither gauge invariance nor custodial $SU(2)$ symmetry
requires this. Such relations arise naturally when one uses an effective
lagrangian in which the breaking $\smgroup\to U(1)_{em}$ is linearly
realized. However, this is a strong assumption -- it is more conservative
to use the nonlinearly realized version of the effective lagrangian in
calculations involving anomalous trilinear gauge boson couplings.

\bigskip
\centerline{\bf Acknowledgments}
\bigskip

This work was supported in part by the Natural Sciences and Engineering
Research Council of Canada, and by FCAR, Qu\'ebec.

\listrefs

\vfill\eject
\centerline{Figures}
\vskip2.5truein
\centerline{Figure~1. Contribution of the CP violating anapole TGC (blob)
to $W$- and $Z$-boson propagators.}
\vskip0.3truecm

\vskip2.5truein
{\vbox{\noindent Figure~2. Diagrams which result in $\phi^3$ and $\phi^9$
couplings once the heavy field $\psi$ is integrated out. The $\psi$ field
is denoted by lines in bold type, while the fine (external) lines represent
the $\phi$ field.}}
\vskip0.3truecm

\vskip1.6truein
\centerline{Figure~3. 3-loop contribution of the $\phi^9$ coupling to the
$\phi^3$ coupling. }
\vskip0.3truecm

\bye